\documentclass[twocolumn,showpacs,preprintnumbers,amsmath,amssymb,floatfix]{revtex4}
\usepackage{graphicx}
\usepackage{dcolumn}
\usepackage{bm}
\begin{document}

\preprint{APS/rapid communication}

\title{Anisotropy in the pion angular distribution of the reaction 
$pp \rightarrow pp \pi^0$ at 400 MeV}

\author{P. Th\"orngren Engblom}
 \email{pia.thorngren@tsl.uu.se}
\author{S. Negasi Keleta}%
\author{F. Cappellaro}
\author{B. H\"oistad}%
\author{M. Jacewicz}%
\author{T.~\ Johansson}%
\author{I. Koch}%
\author{S. Kullander}%
\author{H. Pettersson}%
\author{K. Sch\"onning}%
\author{J. Zloma\'nczuk}%
\affiliation{%
Department of Nuclear and Particle Phsyics, Uppsala University,
Box 535, 751 21 Uppsala, Sweden\\}%
\author{H. Cal\'en}
\author{K. Fransson}
\author{A. Kup\'s\'c}
\author{P. Marciniewski}
\author{M. Wolke}%
\affiliation{%
The Svedberg Laboratory, Uppsala, Sweden}
\author{C. Pauly}
\author{L. Demir{\"o}rs}
\author{W. Scobel}%
\affiliation{%
Hamburg University, Hamburg, Germany}
\author{J. Stepaniak}
\author{J. Zabierowski}
\affiliation{%
Soltan Institute of Nuclear Studies, Warzaw and Lodz, Poland}
\author{M. Bashkanov, H. Clement, O. Khakimova, F. Kren, T. Skorodko}
\affiliation{Physikalisches Institut der Universit\"at T\"ubingen, 
D-72076 T\"ubingen, Germany}
\date{\today}%

\begin{abstract}
The reaction $pp \rightarrow pp \pi^0$ was studied with the WASA detector
at the CELSIUS storage ring. 
The center of mass angular distribution of the $\pi^0$ was obtained 
by detection of the $\gamma$ decay products together with the two 
outgoing protons, and found to be anisotropic with a negative 
second derivative slope, 
in agreement with the theoretical predictions from a microscopic calculation.
\end{abstract}

\maketitle

\section{\label{sec:intro}Introduction}
The first high precision measurements of single neutral pion production
in nucleon-nucleon collisions, using storage ring technology, were done
more than a decade ago. Still, the theoretical interpretation of the
dominant production mechanism remains uncertain.  
The magnitude of the total cross section of the 
$p p \rightarrow  p p \pi^0$ reaction in the threshold region, 
where only angular momenta equal to zero 
are important in the final state, was measured \cite{MEYER92} 
to be about five times larger \cite{MILLER91,NISKANEN92} than what 
was predicted by the theoretical models available at the time.
However, the energy dependence 
was found to be consistent with the widely accepted Koltun and Reitan model 
\cite{KOLTUNREITAN66} based on $s$-wave pion production and rescattering
\cite{MILLER91,NISKANEN92}. The experimental result was confirmed and 
expanded even closer to threshold \cite{BONDAR95} whereas the large 
theoretical activity that was triggered by the new high precision data, 
brought conflicting not yet settled results.

The first successful remedy to fill in the discrepancy between 
experiment and theory was to take into account the exchange of heavy mesons 
\cite{LEERISKA93,HOROWITZ94}. The off-shell pion rescattering (together 
with the Born term) was also suggested to fill in the gap in the cross 
section \cite{HERNANDEZ95}. Both these theories cannot be right unless 
there are some other additional effects. Furthermore, approaches using
chiral perturbation effective field theory (ChPT) reached a different 
conclusion than meson field theory and found the interference between 
the direct term and the pion rescattering to be destructive 
\cite{COHEN96,PARK96}. 
Improved calculations carried out in momentum-space increased the 
rescattering amplitude for the ChPT treatment 
by a factor of three \cite{SATO97}.
Considerable progress has since been made developing the ideas of
\cite{COHEN96}, using an ordering scheme that takes into account the 
large momentum transfer typical for meson production in $NN$ collisions
\cite{BERNARDMEISSNER95,BERNARDMEISSNER99,HANHARTPRL85,HANHARTPRC66}.
Within this scheme it was possible to describe the reaction 
$pp\to d\pi^+$ near threshold \cite{LENSKY05} and a corresponding study 
of the $\pi^0$ production is under way \cite{HANHARTPRIV}. 

A calculation taking into account the exchange of two different heavy 
mesons, pion rescattering and the $P_{11}(1440)$ nucleon resonance reproduced 
the total cross section numbers \cite{KOLK96}.
Relativistic effects were studied in the impulse approximation
in Ref.~\cite{ADAM97}.
The exchange of the mesons $\pi$, $\rho$, $\omega$ and $\sigma$, with the 
nucleon and the $\Delta(1232)$ isobar as intermediate states, 
using a relativistic treatment in a covariant one-boson exchange model 
over an energy range from near threshold to 2 GeV, gave reasonable 
agreement with data \cite{ENGEL96}.
The effect of the resonances $P_{11}(1440)$, $S_{11}(1535)$ and 
$D_{13}(1520)$ together with the impulse and pair diagram terms were 
studied in Ref.\cite{PENA99}

 The possible influence on the differential cross sections due to 
contributions from higher partial waves, including $d$-waves,
was studied experimentally at CELSIUS \cite{ZLOMANCZUK01}.
Angular distributions as well as total cross sections were
recently measured from threshold up to 10 MeV above, by the TOF 
collaboration using an extracted beam \cite{TOF03}. The angular 
distributions were isotropic as expected close to threshold.
The total cross section obtained was about 50 \% larger 
than the IUCF \cite{MEYER92} and CELSIUS data \cite{BONDAR95}. 
The reason for the large deviation compared to the previous storage ring 
experiments was suggested to be due to a significant loss of events in 
the internal experiments, where the very forward going protons escape
down the beampipe undetected. At threshold a strong final state 
interaction could cause the loss of a large number of protons that 
would not be properly accounted for.
\begin{figure}
\includegraphics[width=9.0cm,bb = 0 0 567 384,clip=true]{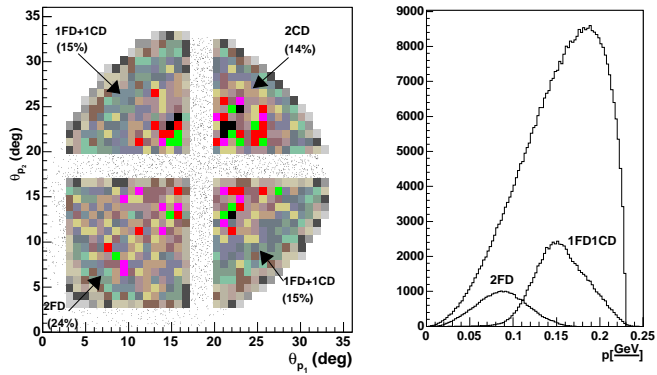}
\caption{\label{fig:acceptance}{\it Left panel :} Scatter plot of the two 
proton polar angles in the laboratory system, obtained from a monte carlo 
simulation of phase space distributed $pp\rightarrow pp \pi^0$ events. 
The geometrical coverage of the WASA detector is shown by the overlaid 
histogram. {\it Right panel :} The relative proton momenta ($p$ from a
simulation according phase space. The labelled curves represent 
different ranges of $p$ corresponding to their detection 
in a full-blown simulation of the detector setup: two forward prongs 
(2FD), one forward and one central (1FD+1CD). The two 
trigger conditions used in the experiment are of the types $2FD$ and $1FD+1CD$ 
(see sec.~\ref{sec:DataAnalysis}).
}
\end{figure}

The data set was drastically increased when the reaction 
$\overrightarrow p \overrightarrow p \rightarrow pp \pi^0$ was measured
at beam energies between 325 and 400 MeV by the PINTEX collaboration at 
IUCF \cite{MEYER98, MEYER99, MEYER00, MEYER01}. 
All possible polarization observables were deduced in the kinematically 
complete experiment and a general formalism was developed from a 
partial wave analysis, in order to obtain an expansion of the 
observables in terms of a complete set of functions mapping the angular
dependence \cite{MEYER01}. 
Thus an analysis method was realized with all the physics information 
contained in the deduced coefficients. However, contributions from $Ds$ and 
$Sd$ to the final state were not taken into account.
The only theoretical model that so 
far has been compared to these data is the microscopic model developed 
by the J\"ulich group \cite{HANHARTPLB444, HANHARTPRC61}. The phenomenology
of the model includes direct production, $s$- and $p$-wave rescattering of 
the pion, pair diagrams and excitation of the $\Delta$(1232). 
Angular momenta up to $L_p,l_q\leq 2$ between the two protons and the pion 
with respect to the NN subsystem, are included.
The same group has recently performed a partial wave analysis using the 
data and the assumptions of \cite{MEYER01} and compared the extracted 
quantities to those of their meson-exchange model \cite{DEEPAK05}. 
Most of the amplitudes are shown to be reproduced fairly well by 
the model, except for the amplitude ${^3P_1}\rightarrow{^3P_0}p$ that deviates
significantly from what is extracted from the data.
For a quantitative assessment of $pp\rightarrow pp\pi^0$ it also turns 
out that the $\Delta$ excitation plays a major role.
For a summary on near threshold meson production experiments 
see \cite{MOSKAL02}. The status of the theoretical field is reviewed in 
\cite{HANHARTrev04}.

In spite of the interest in the reaction $pp\rightarrow pp \pi^0$
during the last 15 years, the reports on the pion angular distributions
at energies up to 400 MeV, suffer from low statistics 
and/or small acceptance. We have measured the 
unpolarized angular cross section of the $\pi^0$ at 400 MeV,
with the aim to resolve some of the ambiguities and complement 
the partial wave analysis based solely on polarized data.

\section{Measurement}
The experiment was done using the WASA 4$\pi$ detector facility \cite{WASA}
situated in the CELSIUS accelerator and storage ring at Uppsala, Sweden. 
A stored circulating proton beam of energy 400 MeV was let to interact 
with a stream of small ($\phi \sim 30$ $\mu$m) frozen hydrogen pellets.
All three outgoing particles from the reaction $pp\rightarrow pp \pi^0$ were 
detected. The protons were fully stopped either both in the Forward 
Detector (FD), or one in the FD and one in the Central Detector (CD). 
The FD consists of a stack of scintillator and wire chamber planes, primarily 
adapted to measure the four momenta of recoiling nuclei.
The CD is constructed for measuring meson decay 
products, and comprises the Scintillating Electromagnetic Calorimeter
(SEC) made up of 1012 CsI detector elmeents, a Plastic Barrel (PB) for charged 
particle detection and a Mini Drift Chamber (MDC) for measuring the momenta 
of charged particles. In the current experiment only the FD and the SEC were 
used for energy measurement and the PB for the rejection of charged particles.
Since the energy and angular resolution for protons was relatively poor in 
the CD, the final analysis was based on the detection of 
the $\pi^0\rightarrow 2\gamma$ decay in the CD.
\begin{figure}
\includegraphics[width=9.0cm]{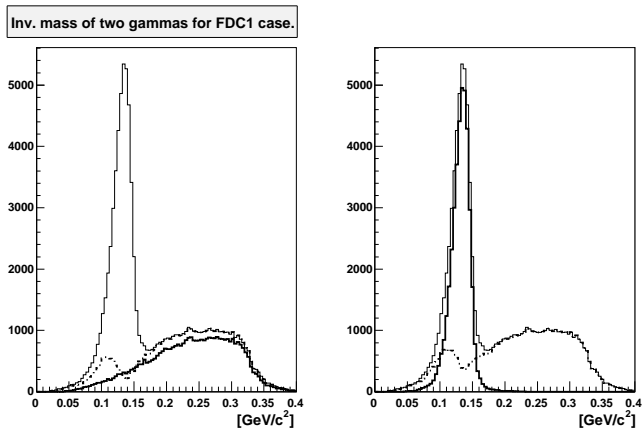}
\caption{\label{fig:InvMass2g} The invariant mass of of the two 
$\gamma$'s selected by the {\it 1FD1CD-type} trigger. 
{\it Left panel :} The cut based on the constraint that the sum of the 
energies of the two $\gamma$'s is within the kinematical limits for 
$\pi^0$ production, is shown by the dashed line. 
The bold line depicts what is cut out based on
the relation between the opening angle and the planarity of 
the two $\gamma$'s, representing largely background from elastic scattering.
{\it Right panel :} The combination of the two cuts
is shown by the dashed line and the final resulting invariant
mass of the two $\gamma$'s, is drawn by the bold line.
}
\end{figure}
\begin{figure}
\includegraphics[width=8.0cm,bb = 0 0 567 384,clip=true]
{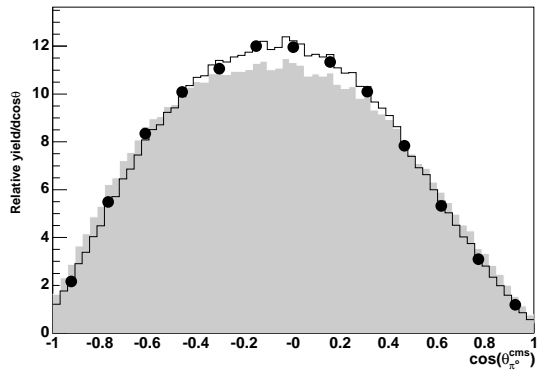}
\caption{\label{fig:expangulardistribution} The experimental 
center-of-mass $\pi^0$ angular distribution, arbitrarily normalized.
The shaded area is the result of a simulated phase space generated 
isotropic distribution of the $pp \rightarrow pp \pi^0$ after passing 
through the detector system. 
The solid line corresponds to the predicted histogrammed values from
a simulation weighted with the microscopic calculation by 
Hanhart et al. \cite{HANHARTPLB444, HANHARTPRC61}. The statistical 
uncertainties are negligible.
}
\end{figure}
\subsection{\label{sec:DataAnalysis}Data Analysis}
The event selection was handled 
using two different sets of criteria based on two different track types, 
which were either both protons detected in the FD ({\it 2FD-type}), or 
one in the FD and one in the CD ({\it 1FD1CD-type}).
The requirements were coincident fast signals from either two hits in one 
scintillator layer of the FD or one in the FD and one hit in the forward 
part of the PB.
These triggers gave an unbiased acceptance of the CD but yielded very high 
countrates, why prescaling was necessary. The main background reactions 
present in the trigger were $pp\rightarrow d\pi^+$ and 
$pp$ elastic scattering. 

The geometrical acceptance for detection of the outgoing protons from the 
reaction  $pp \rightarrow pp \pi^0$ is shown in Fig.~\ref{fig:acceptance}. 
The angular coverage was $3^{\circ}-17^{\circ}$ and $20^{\circ}-155^{\circ}$, 
for the FD and the CD respectively. Since there were no triggers set for the
case when both protons are emitted at $\theta_{lab} > 17^{\circ}$, these events 
escape the current analysis. 
However, the full range of the relative momenta 
of the two protons is covered by the experiment, (c.f. the right panel of 
Fig.~\ref{fig:acceptance}), which is crucial from the physics interpretation
point-of-view.

The basic condition for an accepted event of the $2FD-type$ was particle identification of 
the protons in the FD done by $\Delta E-E$ technique, and the presence in the CD 
of two neutral tracks from the decay of $\pi^0 \rightarrow 2 \gamma$. Additional 
constraints were based on the comparison of the reconstructed polar and 
azimuthal laboratory angles, plus cuts in the center-of-mass energy, with 
respect to the missing mass of the two protons and the invariant mass of the 
two $\gamma$'s from the $\pi^0$ decay.  
The conditions applied were: $|\theta_{Mx} - \theta_{IM}| < 15^{\circ}$,
$|\phi_{Mx} - \phi_{IM}| < 15^{\circ}$ and $|{E_{Mx} - E_{IM}}| < 30$ MeV, 
where $Mx$ is the missing mass of the two protons and $IM$ the invariant mass
of the two $\gamma$'s.
The consistency of the $\pi^0$ angle reconstructed from the missing mass 
and the invariant mass, respectively, was investigated.
The two approaches agreed and thus the conclusion was drawn that analysis 
of the {\it 1FD1CD-type} prongs could be done using CD information only.

The selection of event candidates for the type of 
events with one forward and one central prong ($1FD1CD-type$) was done by particle 
identification of the FD proton. Furthermore, for the two $\gamma$'s 
a cut was based on the relation between the opening and planarity angles.
An additional constraint applied was that the sum of the energies of the 
two $\gamma$'s was within the kinematical limits for 
$\pi^0$ production, i.e. $135 < \Sigma E_{\gamma} < 238$ MeV,
 The $\pi^0$ peak obtained from the invariant mass 
of the two $\gamma$'s, before and after track requirements are fulfilled,
is shown in Fig.~\ref{fig:InvMass2g}. 

The two sets of selected $pp\pi^0$ events were weighted together
according to their relative trigger prescaling factor. 
The experimental angular distribution of the $\pi^0$ in the center-of-mass, 
uncorrected for acceptance, is seen in Fig. \ref{fig:expangulardistribution}. 
Displayed are also simulations using either isotropically distributed events 
according to phase space or events weighted with the theoretical calculation 
of the J\"ulich model by Hanhart et al. \cite{HANHARTPLB444, HANHARTPRC61}. 
More details on the data reduction 
procedure can be found in \cite{SAMSONlic04}.

\begin{figure}
\includegraphics[width=8.0cm,bb=50 180 650 700,clip=true]{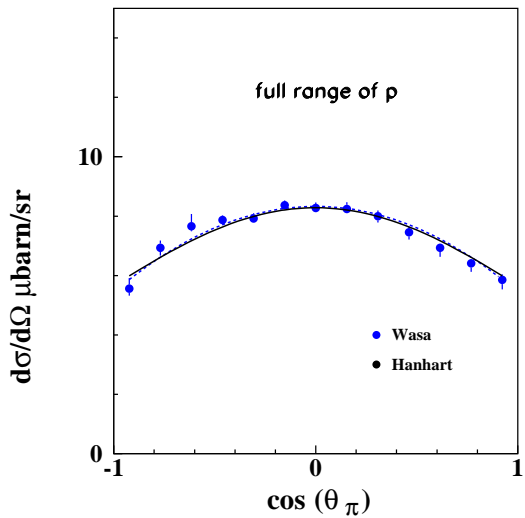}
\caption{\label{fig:acceptancecorrected}Acceptance corrected 
center-of-mass $\pi^0$ angular distribution, normalized to 
$\sigma_{tot}=92.3\pm7.2$ $\mu$barn \cite{ZLOMANCZUK01}. 
The uncertainites are dominated by systematic effects. The solid line corresponds to 
the microscopic calculation by Hanhart et al. \cite{HANHARTPLB444, HANHARTPRC61}. 
The dashed line represents a fit of the dependence on $\cos^2\theta_{\pi}$,
taking only the statistical errors into account.
}
\end{figure}
\begin{figure}
\includegraphics[width=8.0cm,bb=50 180 650 700,clip=true]{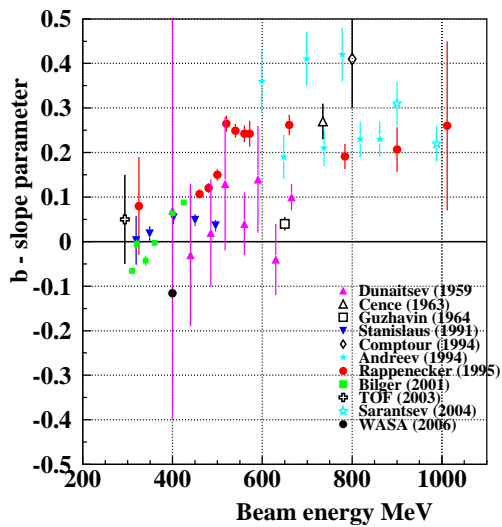}
\caption{\label{fig:bpar} A compilation of measurements of the slope
parameter below 1 GeV is shown. The definition of $b$ differs by a factor of three in 
\cite{ZLOMANCZUK01} why the values given by that reference have been divided 
by three for consistency. (The large error bar at 400 MeV is from \cite{DUNAITSEV59}).
}
\end{figure}
\begin{figure}
\includegraphics[width=8.0cm,bb=50 180 650 700,clip=true]{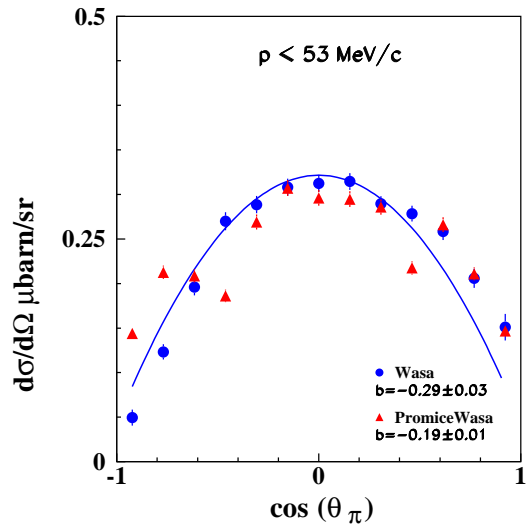}
\caption{\label{fig:plt53} Acceptance corrected 
center-of-mass $\pi^0$ angular distribution, for $p < 53$ MeV/c, compared and 
normalized to \cite{ZLOMANCZUK01}. 
The solid line represents a fit of the WASA data points 
to the dependence on $\cos^2\theta_{\pi}$, 
taking only the statistical errors into account.
}
\end{figure}
\section{Results}
The acceptance corrected $\pi^0$ angular distribution is  
shown together with the prediction by the J\"ulich model in
Fig. \ref{fig:acceptancecorrected}. The experimental data points and 
the theoretical curve are normalized to $\sigma_{tot}=93\pm7.2$ $\mu$barn 
from \cite{ZLOMANCZUK01}. 
The systematic uncertainties dominate, primarily emanating from the acceptance 
varying with the central detector's geometry. In order to obtain an estimate of 
the magnitude of this error the outmost layers in the forward and the backward parts 
respectively, were excluded in the analysis.  

In previous experimental reports 
\cite{DUNAITSEV59,CENCE63,STANISLAUS91,COMPTOUR94,ANDREEV94,RAPPENECKER95,SARANTSEV04,TOF03} 
concerning the angular distribution a slope parameter $b$ was defined
according $\frac{d\sigma}{d\Omega}\propto \frac{1}{3}+b\cos^2\theta_{\pi}$, see 
Fig.\ref{fig:bpar} for a compilation of measurements below 1 GeV. There is a large
spread of the values, probably mainly due to the varying coverage of $p$ in the 
measurements. One recent experiment \cite{ZLOMANCZUK01} yield a negative 
$b$ up to 360 MeV beam energy but at 400 MeV the slope was reported to be positive
in discrepancy with the present result $b=-0.116\pm 0.010$.
However the acceptance of \cite{ZLOMANCZUK01} was limited with respect to $p$, 
with a coverage similar to the $2FD-type$ case, and a model dependence was 
introduced by extrapolating into unmeasured regions of phase space. 
It should be noted that events with {\it both} proton angles larger than $17^\circ$ 
were not detected within the current acceptance, see Fig. \ref{fig:acceptance}. 
There are indications that the inclusion of such events would slightly flatten
the distribution \cite{CLEMENTPRIV}. 
\setcounter{table}{0}
\begin{table}
\caption{Correspondence between $H_1^{00}+I$ and b}
\begin{ruledtabular} 
\begin{tabular}{l|ccc}\label{table:H2b}
 & WASA & \cite{ZLOMANCZUK01} & \cite{MEYER01}(pol) \\
\hline
$H_1^{00}+I$ & $-0.131\pm0.012$ & $0.055\pm0.007$ & $0.084\pm0.053$ \\
$b$  & $-0.116\pm0.010$  & $ 0.063\pm0.003$\footnotemark[1]  & $ 0.09\pm0.18$ \\
\end{tabular}
\end{ruledtabular}
\footnotemark[1]{Published value is divided by three}
\end{table}
Whereas the $\pi^0$ angular distribution integrated over all $p$ displays
discrepancies among the different experiments, a selection of $S$-wave protons
might shed some light. At very low relative momenta between the protons,
$p < 53$ MeV/c, the two CELSIUS measurements agree at least qualitatively, 
see Fig. \ref{fig:plt53}. At 800 MeV beam energy using the same cut in $p$,
an even larger negative second derivative was found \cite{DYMOV05}, 
which was also predicted by a phenomenological calculation\cite{NISKANEN06}.

\begin{figure}
\includegraphics[width=7.0cm,bb=50 180 650 700,clip=true]{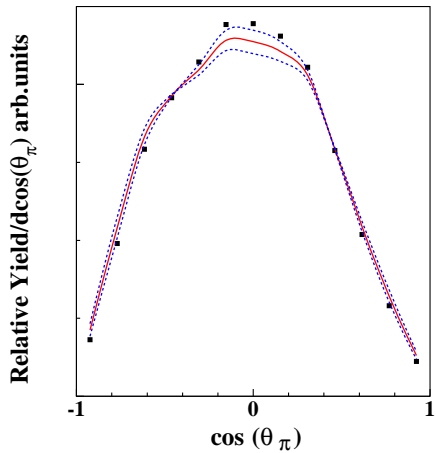}
\caption{\label{fig:sampling}Using the sampling method \cite{KUROS} 
for direct comparison between the experimental data and the expansion 
developed in \cite{MEYER01}. The dashed lines  
correspond to the uncertainty ($1\sigma$) in the determination of the coefficient 
combination $H_1^{00}+I$.
}
\end{figure}

A direct comparison between the present experiment and the expansion deduced from
the double polarization data \cite{MEYER01} can be done by using the 
{\it sampling method} \cite{KUROS}, that allows to integrate the prediction of 
a theoretical model over the experimentally accessible phase space region, 
see Fig.~\ref{fig:sampling}.
The coefficient combination $H_1^{00}+I$ and the dependency on the 
$\cos^2(\theta _\pi)$-term (Eq. 11 \cite{MEYER01}), have been taken into account, 
all other variables are ignored. See Table \ref{table:H2b} for the correspondence
between the slope parameter $b$ and $H_1^{00}+I$.

The prospect of improving the accuracy of certain coefficients, as well
as pinning down the only remaining undetermined coefficient, $H_3^{00}$,
using the present experimental event sample, is now a highly feasible 
plan for the immediate future. Thus another advancement has been made towards 
a complete characterization of the amplitudes of the fundamental reaction 
$pp\rightarrow pp\pi^0$ at low energy.
With all the data available, both polarized and 
unpolarized, in conjunction with the {\it sampling method}, we anticipate
that all amplitudes can be determined individually. For the future development 
of ChPT, detailed realistic constraints will be supplied.
\begin{acknowledgments}
We wish to acknowledge the support of the Swedish Research Council with
Grant Nos. Dnr. 629-2001-3868 and Dnr. 40313601, and the G\"oran 
Gustafsson Foundation Grant No. 01:95. The hard work of the CELSIUS crew
is greatly appreciated. One of the authors, P.T.E., is grateful for 
clarifying discussions concerning the reaction $pp \rightarrow pp \pi ^0$ 
with Prof. H.-O. Meyer, Dr. C. Hanhart, Prof. A. Johansson
and Prof. C. Wilkin.
\end{acknowledgments}
\bibliography{pppi0}
\end{document}